\documentclass{ws-ijmpd}

\begin{document}

\title{Weyl transformation: a dynamical degree of freedom in the light of Dirac's Large Number Hypothesis}

\author{PRASENJIT PAUL}
\address{Department of Physics, Government College of Engineering and Ceramic Technology, Kolkata 700 010, West Bengal, India \\Department of Physics, Indian Institute of Engineering Science and Technology, Howrah \\711 103, West Bengal, India \\prasenjit071083@gmail.com}

\author{RIKPRATIK SENGUPTA}
\address{Department of Physics, Government College of Engineering and Ceramic Technology, Kolkata 700 010, West Bengal, India \\ rikpratik.sengupta@gmail.com}

\author{SAIBAL RAY}
\address{Department of Physics, Government College of Engineering and Ceramic Technology, Kolkata 700 010, West Bengal, India \\ saibal@associates.iucaa.in}

\maketitle

\begin{history}
\received{Day Month Year} \revised{Day Month Year} \comby{Managing Editor}
\end{history}

\begin{abstract}
	 In Einstein's Field Equation(EFE) the geometry of the space-time is connected with the matter distribution. The geometry or the gravitational sector deals with classical macroscopic objects involving gravitational units while the matter sector can be better described by quantum theory involving atomic units. It has been argued by Bisabr that there exists an epoch-dependent conversion factor between these two unit systems present in two different conformal frames,i.e. the conformal factor is epoch dependent. We argue that the conformal transformation is a dynamical degree of freedom describing it's possible relevance in inflation in context to the graceful exit problem, dynamics of the cosmological constant $\Lambda$ and justify the argument in the light of consequences of Dirac's Large Number hypothesis(LNH). 

\end{abstract}

\section{Introduction}

Following the argument by Bisabr~\cite{Bisabr},we consider the geometry and matter parts of the EFE,or simply the gravitational and the matter sectors to be in two different conformal frames related by an epoch-dependent or time dependent transformation factor. The main logic behind this is that gravitational systems being large are described in the classical regime while matter sector is described on a microscopic level in the quantum regime. This raises a question whether the equivalence principle(EP) holds true in all epochs or phases of the evolution of the Universe.  

Though we have experimental verification~\cite{Will1993,Damour1996,Damour2000,Will2005} of EP,there remains a question about its consistency in the regime where the laws of quantum physics hold true~\cite{Davies1982,Cho1997}. There exist some interacting scalar fields like the dilaton field occuring in string theory, the consideration of whose dynamics does not allow EP to hold true~\cite{Damour2001,Damour2012}. If the fields have a long range effect then the EP violation may have direct observational consequences. This fact is also true in a semiclassical treatment invoking the uncertainty principle which will not allow freely falling objects to follow a specific trajectory or even in a classical treatment~\cite{Damour1990}.
 
So far no scalar field has been observed in any experiment. A possible instance of this was provided by Khoury and Weltman~\cite{Khoury2004a,Khoury2004b} with the introduction of a chameleon field. The time dependence of the fine-structure constant along with the idea of dark energy led to the concept of a chameleon field whose matter-coupling has to be tuned to extremely fine values to avoid violation of EP. At large scales in the regions where density is low, the chameleon field may have it's observable effects through the difference of effective Newton's constant in space than on earth and hence violation of EP.

 The paper has been organized as follows : In Sec. 2 we make a brief review on Dirac's Large Number Hypothesis (LNH). In Sec. 3, a model has been discussed in which the matter and gravitational sectors of the EFE are assumed to be present in two different conformal frames (CFs). We argue in the light of LNH that the conformal factor happens to be a dynamical degree of freedom. This fact is crucial providing possible explanation of the smallness of the cosmological constant in the late universe and also resolves the graceful-exit problem in the early universe as discussed in Sec. 4.  In Sec. 5, we provide concluding remarks remarking on the justification of our argument.

\section{A brief review on Dirac's LNH}
 Though a long time has passed since the introduction of the LNH by Dirac, it is still significant. On the contrary,it has become more significant after the discovery of accelerating Universe from SN Ia observations\cite{Perlmutter1998,Riess1998} which advocates the idea of a variable epoch-dependent cosmological constant $\Lambda$ as a possible candidate for explaining the repulsive force causing the late-time universe to accelerate. Compatibility with the rest of physics demands that $\Lambda$,which is effectively the energy of the vacuum should be gradually decreasing with time. If we cosider $\Lambda$ as the dark energy or the repulsive effect,then the presently observed small ($\approx 10^{-35}s^{-2}$) value of $\Lambda$ may be possibly explained by the fact that it has decreased gradually from large initial value to this extremely tiny value at present~\cite{Carmeli2002a,Carmeli2002b}. This time-dependent nature possibly exhibited by $\Lambda$ is analogous to the main spirit of the idea of LNH which advocates a time-dependent dynamic gravitational constant $G$ as suggested by Dirac~\cite{Peebles2003}. Both the cosmological constant and gravitational constant are not really constants but parameters that have decreased gradually from high initial to late time small values in course of time-evolution of the universe.
 
 A causal connection between the macroscopic and microscopic worlds can be realized from LNH. Dirac~\cite{Dirac1937,Dirac1938} pointed out that the ratio of electric ($e^2/4 \pi \epsilon_0 r^2$) and gravitational ($G m_p m_e/r^2$) forces between a proton and an electron in the hydrogen atom is quite a large number,being of the order of $10^{40}$, i.e.
 \begin{eqnarray}
 \frac{e^2}{4 \pi \epsilon_0 G m_p m_e}\approx 10^{40}, \label{eq1}
 \end{eqnarray}
 where $e$ is the electronic charge, $m_p$ is the mass of a proton and $\epsilon_0$ is the permittivity of vacuum.
 
 Again, the ratio of the then age of the Universe ($2 \times 10^9$) and the atomic unit of time ($e^2/4 \pi \epsilon_0 m_e c^3$) is also nearly of the same size, i.e. 
 \begin{eqnarray}
 \frac{4 \pi \epsilon_0 m_e c^3}{e^2}\approx 10^{40}. \label{eq2}
 \end{eqnarray}
 
 Dirac claimed that the two quantities in the left hand side of Eqs. (\ref{eq1}) and (\ref{eq2}) must therefore be equal, i.e.
 \begin{eqnarray}
 \frac{e^2}{4 \pi \epsilon_0 G m_p m_e}\simeq \frac{4 \pi \epsilon_0 m_e c^3}{e^2}. \label{eq3}
 \end{eqnarray}
 
Relying on Eq. (\ref{eq3}), Dirac established, based on the argument of causal connection, the very slow variation of the fundamental constants. This is the theme of Dirac's LNH in a nutshell (an exhaustive review on this can be obtained in the work by Ray et al.~\cite{Ray2019}). Thus, the hypothesis claims that that as the atomic parameters are not subject to time-evolution,so the gravitational constant or rather parameter exhibits an inverse time variation expressed in atomic units, i.e.
\begin{eqnarray}
G \propto \frac{1}{t}. \label{eq4}
\end{eqnarray}
 
It is argued that LNH requires the continuous creation of matter in the universe~\cite{Ray2019}. There are two possible ways of matter creation: (i) additive creation and (ii) multiplicative creation. In the former,matter forms even in intergalactic space, whereas in the later,matter forms only at specific places where there is previous presence of matter and the rate of matter formation is directly proportional to  pre-existing matter quantity.

\section{Weyl transformation as a mathematical tool in unit transformation}

\subsection{Weyl transformations}
A line element associated with a metric $g_{\alpha\beta}$ is given by $ds^2=g_{\alpha\beta}dx^{\alpha}dx^{\beta}$ and hence a conformal transformation can be defined by~\cite{Hawking1973,Wald1984}
\begin{equation}
\bar{g}_{\alpha\beta} =e^{-2\xi}  g_{\alpha\beta}, \label{eq5}
\end{equation}
where $\xi=\xi(x^{\alpha})$ depends on spacetime which is a smooth and dimensionless function. 

The transformation relation of the above mentioned line element can be written as
\begin{equation}
ds\rightarrow d\bar{s}=e^{-\xi}ds. \label{eq6}
\end{equation}

This change in spacetime intervals through the conformal transformation (CT) can be regarded as a unit transformation. Usually a global unit transformation is represented by a conversion factor which is constant, e.g. transformation from mks system to cgs system is a global unit transformation. But in case of CT we get a local unit transformation~\cite{Dicke1962,Bekenstein1980} and in this case transformation takes place for all quantities in accordance with their dimensions. Now, if we consider $\phi$ as a standard model (SM) field then under CT we have
\begin{equation}
\bar{\phi} = e^{\alpha\xi} \phi, \label{eq7}
\end{equation}
where $\alpha$ is a number which determines the type of the SM field, $\phi$. If we consider scalar field then $\alpha=1$
whereas for Dirac fields $\alpha$=3/2~\cite{Bekenstein1980}.

\subsection{Relation of fundamental constants between different unit systems}
Two basic unit systems have been considered over the years in the fields of Quantum Mechanics and General Relativity. In the former field,the atomic system of units (asu) is largely used, described by constants, such as the masses ($m$), charges ($e$) of elementary type particles, $c$ and $\hbar$. Whereas,in the latter field the gravitational system of units or gsu consisting of the masses ($M$), the gravitational constant, $G$ and macroscopic object sizes ($R$) finds extensive use. Every unit systems comprise of a whole set of constants in a manner that construction of the units in terms of time, space and also mass for one elements do not depend on elements corresponding to the other set. Usually it has been considered that switching between two unit systems can be done via a constant transformation factor. Its implication is the fact that any one of such unit systems is basically a constant product of other unit systems. Now if one designate the spacetime intervals for the gravitational units and atomic units as $ds_{G}$ and $ds_{A}$ respectively, then $ds_{G}=\gamma~ds_A$ where $\gamma$ is actually a constant factor. The transformation factor $\gamma$ can be represented by means of the so called fundamental constants like $e$, $G$, $\hbar$, $c$ etc. So whenever it is assumed that $\gamma$ being a constant, it can be taken for granted that these very fundamental physical quantities are also constants. However, one can utilise one unit system conveniently and it is irrespective of any dynamical meaning. That means one can arbitrarily utilise asu to analyse the development of the Universe and may also utilise gsu to analyse dynamical behaviour of any particle of elementary type.

From Dirac's LNH~\cite{Dirac1973} one can draw few predictions, which contradict the fact that $\gamma$ is constant. Dirac assumed that the spacetime interval for a gravitational theory, $ds_{G}$ is different from that in case of the atomic theory, $ds_{A}$ measured by an atomic apparatus. 

Einstein’s GR does not accomodate a dynamic gravitational constant $G$ due to the vanishing divergence  of the Einstein tensor $G_{\alpha\beta}$ which is essential for the energy conservation law $T_{; \beta}^{\alpha\beta}=0$ to hold true. As a way out, Dirac treated two metrics where the Einstein metric leads to the equations of motion and decribes the unchanged classical sector. Another one is the atomic metric involving atomic quantities computing times and distances by equipment used in laboratory~\cite{Faulkner1976}. Threfore, the intervals $ds_A$ and  $ds_G$ between two events in asu and gsu respectively will differ. So as far as gsu or asu is converted to other unit system the equations written in them cannot be used at a time~\cite{Rogachev2006}. Here in both metrics the speed of light has been taken to be unity. If the Solar system is considered,it has been shown that~\cite{Dirac1974} the additive and multiplicative creation mechanisms result in contrasting consequences as the relation between the atomic and the Einstein metrics is not identical for the two mechanisms.

To explain the fact that the ratio of the metrics has an epoch dependence, Dirac developed a theory with variable $G$. Canuto et al.~\cite{Canuto1977a,Canuto1978} develop a  gravitational theory with scale-covariance in later years and studied the consequences of this theory in astrophysical manner. In such type of consideration, the quantities belonging to the atomic theory e.g. $e$, $m$ and $\hbar$ remain constants in asu whereas $G$, $M$ and $R$ may vary and vie versa. So, it can be stated that the progressive or dynamical difference of these unit systems circles around the argument of dynamic nature of physical constants. As $\gamma$ can be expressed in terms of the fundamental physical constants as mentioned already, any symptom of the changes of these constants indicate the corresponding epoch dependent changes of $\gamma$. Though till now no observational result show these type of variations, any future prediction of these variations points out that gsu and asu obviously need to be taken as dynamically different.

\subsection{Scale-covariance}
As mentioned earlier, $G$ being a constant in the Einstein equation one can not lightly take it as variable. So, in the light of Dirac’s LNH, some changes of EFE are required. If we take $G$ as a variable, then there will be violation of energy conservation law~\cite{Canuto1977b,Wesson1981}. So, the field equations and conservation laws need to be modified to study the consequence of varying $G$. There exist a scale-covariant theory~\cite{Canuto1977b,Canuto1977c} in which a gauge function $\gamma$ is selected. The main aspect of scale-covariant theory  has the central theme of invariance of physical laws for alternative selection of gauge function. The usual Einstein equation can be written as
\begin{eqnarray}
\bar{G}_{\alpha\beta} = -8\pi \bar{T}_{\alpha\beta}+ \Lambda \bar{g}_{\alpha\beta}, \label{eq8}
\end{eqnarray}
where the corresponding line element is given by
\begin{eqnarray}
{\bar{ds}}^2 = {\bar g}_{\alpha\beta} dx^{\alpha} dx^{\beta}. \label{eq9}
\end{eqnarray}
Considering the transformation
\begin{eqnarray}
ds = \gamma^{-1}(x)\bar {ds}, \label{eq10}
\end{eqnarray}
and by taking modified Ricci tensors~\cite{Eisenhart1926} the field Eq. (\ref{eq8}), in general units gets the modified form as follows~\cite{Canuto1977b}
\begin{eqnarray}
G_{\alpha\beta}+ 2\frac{\gamma_{\alpha ; \beta}}{\gamma}- 4\frac{\gamma_\alpha \gamma_\beta}{\gamma^2}- g_{\alpha\beta}\left(2\frac{\gamma^{\mu}_{;\mu}}{\gamma}- \frac{\gamma^{\mu} \gamma_{\mu}}{\gamma^2}\right) = -8\pi T_{\alpha\beta}+ \Lambda g_{\alpha\beta}, \label{eq11}
\end{eqnarray}
where $T_{\alpha\beta}$ and $\Lambda$ are the energy-momentum tensor and the cosmological constant respectively. This $\Lambda$ is related to $\bar{\Lambda}$ as
\begin{eqnarray}
\Lambda = \gamma^2 \bar{\Lambda}.  \label{eq12}
\end{eqnarray}

Hence the modified energy conservation law can be written as
\begin{eqnarray}
\dot{\rho}+ (\rho+p)u^{\alpha}_{; \alpha} =-\rho \left(\frac{\dot G}{G}+ \frac{\dot{\gamma}}{\gamma}\right)- 3p\frac{\dot{\gamma}}{\gamma}.  \label{eq13}
\end{eqnarray}

The Friedmann-Robertson-Walker metric 
\begin{eqnarray}
ds^2 = dt^2- a^2(t)\left(\frac{dr^2}{1-kr^2}+ r^2 d\theta^2+ r^2 \sin^2\
\theta d\phi^2\right),  \label{eq14}
\end{eqnarray}
in the modified form becomes
\begin{eqnarray}
\left(\frac{\dot a}{a}+ \frac{\dot{\gamma}}{\gamma}\right)^2+ \frac{k}{r^2} = \frac{8\pi G\rho}{3}+ \frac{\Lambda}{3}, \label{eq15}
\end{eqnarray}

\begin{eqnarray}
\frac{\ddot a}{a}+ \frac{\ddot{\gamma}}{\gamma}+ \frac{\dot{\gamma}}{\gamma}\frac{\dot a}{a}- \frac{\dot{\gamma}^2}{\gamma^2} = -\frac{4\pi G}{3}(3p+\rho)+ \frac{\Lambda}{3}. \label{eq16}
\end{eqnarray}

We can write the energy conservation laws as
\begin{eqnarray}
\dot{\rho}+ 3\frac{\dot a}{a}(\rho+p) = -\frac{\rho}{G \gamma} \frac{d}{dt}(G \gamma)- 3p \frac{\dot{\gamma}}{\gamma}. \label{eq17}
\end{eqnarray}

In case of the scale-invariant (scale-covariant) theory~\cite{Canuto1977a,Canuto1977b}, the cosmological constant $\Lambda$ changes as $\gamma^2$ and it is not a constant. As the evaluation of $\gamma(t)$ provides us the opportunity to collate the theory with observational results, therefore its determination is very vital for scale-covariant theory. It is hardly possible to evaluate $\gamma(t)$ within the theory since it provides us the freedom in selecting the system of units. Hence one cannot construct any dynamical equation for evaluating $\gamma(t)$ and so dissimulation of external constraint is compulsory in this case. Considering relation with the cosmological parameter and gauge fields one can suggest that $\gamma$ is actually inversely proportional to $t$. Again, adapting Dirac's LNH, one can in lieu of show that $\gamma = {t_0}/t$, where $t_0$ represents the present age of the Universe.~\cite{Canuto1977b,Canuto1977c,Canuto1977d}.

Now we shall inspect the consequences of the following guess
\begin{equation}
ds_G=\gamma^{-1}(t)~ds_A, \label{eq18}
\end{equation}
where $\gamma(t)$ is an epoch function. Comparing (\ref{eq18}) with (\ref{eq6}) and (\ref{eq10}) one can show that it represents actually a conformal transformation with $\gamma^{-1}$ as a conformal factor. As the ratio of a particular spacetime interval in gsu and asu increases in an expanding Universe, therefore (\ref{eq18}) shows that $\gamma(t)$ must decrease with time as assumed earlier in this Section. So changing unit systems involves a dynamical content. 

However, one may argue that this fact brings about an ambiguity when one makes a comparison of a physical quantity in quantum physics
with its analogous value in gravitational theory. As for example, if we consider the time-dependent cosmological constant, i.e. $\Lambda(t)$, then its theoretical evaluation in particle physics are done in the atomic unit and it does conflict with the observations which are  made in the gravitational unit. So if we do not consider the dynamics of $\gamma(t)$ the cosmological constant problem arises. On the other hand, if we consider gsu and asu are identical up to a constant conversion factor then a huge difference occur between observations and theoretical estimated values. However, this difference does not arise when the unit transformation is considered to be dynamical.

\section{Inflationary solutions for quadratic and cubic potentials}
In this Section we shall use Weyl transformation to obtain inflationary cosmological solutions considering the potential of the inflaton field to have quadratic and cubic nature as different cases. Weyl transformations may be visualized as a tool for mapping the equations of motion of a physical system to an equivalent set of simpler mathematical equations. The basic assumption usually taken to obtain solutions is that different forms of matter are coupled to gravity described by same background geometry in the form of a unique metric. However, as said earlier, there may be phases in the evolution of the Universe where the EP does not hold true. In that case there will be two metrics - one for the gravitational sector and another for the matter sector, pertaining to two different conformal frames, i.e. the metrics are connected by a CT (as specified 
by the works~\cite{Hawking1973,Wald1984}) to have the form as Eq. (\ref{eq5}) and the inflaton transforms as given by Eq. (\ref{eq7}). 

The inverse metric and the determinant then transform under CT as
\begin{equation}
\bar{g}^{\alpha\beta} =e^{2\xi}  g^{\alpha\beta}, \label{eq19}
\end{equation}

\begin{equation}
\sqrt{-\bar{g}} = e^{-4\xi} \sqrt{-g}. \label{eq20}
\end{equation} 

The Lagrangian density for the inflaton, which must be invariant under the CT is given as
\begin{equation}
{\mathcal{L}}(\bar{g}_{\alpha\beta}, \bar{\psi})=\frac{1}{2}\bar{g}^{\alpha\beta}\nabla_{\alpha}\bar{\psi}\nabla_{\beta}\bar{\psi}+V(\bar{\psi}), \label{eq21}
\end{equation}
with the full action of the gravity-matter system given by
\begin{equation}
S=\frac{1}{2k} \int d^{4}x \sqrt{-g} {\mathcal{R}} -\int d^{4}x \sqrt{-\bar{g}} {\mathcal{L}}(\bar{g}_{\alpha\beta} \bar{\psi}). \label{eq22}
\end{equation}

The condition for slow-roll approximation in this case may be written down as $\{(\partial \psi)^2, \Box \psi \} << V(e^{\xi}\psi)e^{-4\xi}$.
Application of this condition gives the final form of the action as
\begin{equation}
 S=\frac{1}{2} \int d^{4}x \sqrt{-g} \left\{\frac{1}{k}{\mathcal{R}}-\psi^2 g^{\alpha\beta}\partial_{\alpha} \xi \partial_{\beta}\xi-V(e^{\xi}\psi)e^{-4\xi}\right\}.  \label{eq23}
\end{equation}
   
Now we shall consider two different models of inflation for quadratic and cubic potentials of the inflation field.

\subsection{Model for quadratic inflation field}
The potential of the inflaton field has the form~\cite{Linde1983}
\begin{equation}
V(\bar{\psi})= \nu m_p^4\left(\frac{\bar{\psi}}{m_p}\right)^2, \label{eq24}
\end{equation} 
where $\nu$ and $p$ are constant quantities with $\nu<<1$ and ${m_p}^2=\frac{1}{G}$, $m_p$ denoting the Planck mass.

The slow-roll parameters are given by
\begin{equation}
\epsilon=\frac{4}{2\lambda^2}~,~ \eta=\frac{2}{\lambda^2}, \label{eq25}
\end{equation}
where  $\lambda \equiv \psi/m_p$ and during slow-roll inflation we can write $\lambda >> 1$.

The field equations can be computed to be
\begin{equation}
3 \left(\frac{\dot{a}}{a}\right)^2=\frac{1}{2}k \psi^2\left\{\dot{\xi}^2+\nu m_p^{2} e^{-2\xi}\right\}, \label{eq26}
\end{equation}

\begin{equation}
\ddot{\xi}+3H \dot{\xi}-\nu m_p^{2}e^{-2\xi}=0. \label{eq27}
\end{equation} 

The solution to these field equations are obtained as 
\begin{equation}
a(t) \sim t^q, \label{eq28}
\end{equation}

\begin{equation}
\xi(t)=\xi_T +\ln {\left(\frac{t}{t_T}\right)}, \label{eq29}
\end{equation} 
where
\begin{equation}
 q= 4\pi \lambda^2,~t_T^2=\left[\frac{12\pi\lambda^2-1}{\nu m_p^2 e^{-2\xi_T}}\right], \label{eq30}
\end{equation}
where  $\xi_T$ is a dimensionless constant. The effective cosmological constant can be written as 

\begin{equation}
\Lambda_{eff} \sim 4\pi\nu^2m_p^2\gamma^2e^{-2\xi}.
\end{equation}

\subsection{Model for cubic inflation field}
The potential of the inflaton field has the form~\cite{Linde1983}
\begin{equation}
V(\bar{\psi})= \nu m_p^4\left(\frac{\bar{\psi}}{m_p}\right)^3.
\end{equation} 
where $\nu$ and $p$ are constant quantities with $\nu<<1$ and ${m_p}^2=\frac{1}{G}$, $m_p$ denoting the Planck mass.

The slow-roll parameters are given by
\begin{equation}
\epsilon=\frac{9}{2\lambda^2}~,~ \eta=\frac{6}{\lambda^2}, \label{eq31}
\end{equation}
where again $\lambda \equiv \psi/m_p$ and during slow-roll inflation we can write $\lambda >> 1$.

The field equations in this case can similarly be computed as
\begin{equation}
3 \left(\frac{\dot{a}}{a}\right)^2=\frac{1}{2}k \psi^2\left\{\dot{\xi}^2+\nu m_p \psi e^{-\xi}\right\}, \label{eq32}
\end{equation}

\begin{equation}
\ddot{\xi}+3H \dot{\xi}-\frac{1}{2}\nu m_p\psi e^{-\xi}=0. \label{eq33}
\end{equation} 

The solution to these field equations are obtained as 
\begin{equation}
a(t) \sim t^q, \label{eq34}
\end{equation}

\begin{equation}
\xi(t)=\xi_T +2\ln {\left(\frac{t}{t_T}\right)}, \label{eq35}
\end{equation} 
where
\begin{equation}
q= 16\pi \lambda^2,~t_T^2=\left[\frac{4(48\pi\lambda^2-1)}{\nu \lambda m_p^2 e^{-\xi_T}}\right], \label{eq36}
\end{equation}
where  $\xi_T$ is a dimensionless constant. The effective cosmological constant for the cubic potential is given by

\begin{equation}
\Lambda_{eff} \sim 4\pi\nu^2m_p^2\gamma^3e^{-\xi}.
\end{equation}

We note that in both the models a power-law type of solution is obtained. Moreover, if the inflaton has a quadratic potential then the conformal field equation is found to be field-independent. The constant $\xi_T$ can be interpreted physically as the value of $\xi$ at which the inflation ends.

For both the potentials there is a decay in the effective cosmological constant arising out of the conformal factor and in both the cases it turns out that $\Lambda_{eff} \sim t^{-2}$. This provides a possible solution to the cosmological constant problem for a flat universe. 

Also here, the energy density of the inflaton field,rather than remaining at a fixed value as in the case of an exponential inflation,decays exponentially which makes the potential energy of the inflaton to be of the order of the kinetic energy,thus leading to the violation of the slow-roll approximation and termination of the inflationary mechanism. Thus we have a possible solution to the graceful-exit problem.

\section{Conclusion}
About seven decades ago, Dirac mentioned the possibility of variation of the fundamental constants in the cosmological scenario. Over the years various new ideas have been developed building upon LNH. We have shown that the idea of time varying nature of macroscopic constants finds relevance in explaining the smallness of the cosmological constant. We see that considering inflationary solutions with quadratic and cubic potentials for the inflaton, where the inflaton and gravitational metric are in two different conformal frames, we obtain a conformal factor which is dynamic,varying logarithmically with time. This justifies the Weyl transformation as a dynamical degree of freedom which results in the decay of the potential of the inflaton helping terminate inflation providing a possible solution to the graceful exit problem. Also,the application of Weyl transformation to slow-roll inflation provides a dynamical reduction mechanism for the cosmological constant which works during evolution of the Universe which may have significant imprint on the small value of $\Lambda$ in the present universe. This is in confirmity with the time decreasing(dynamic) nature of $\Lambda$ as predicted by Dirac's LNH.

\section*{Acknowledgments} SR is thankful to the authority of Inter-University Centre for Astronomy and Astrophysics, Pune,
India for providing him Visiting Associateship under which a part of this work was carried out.

\end{document}